\documentclass[aps,prc,reprint,twocolumn,superscriptaddress,nofootinbib]{revtex4-2}
\usepackage{graphicx,epsfig}
\usepackage{amsfonts}
\usepackage{amsmath}
\usepackage{amssymb}
\usepackage{braket}
\usepackage{multirow}
\usepackage{commath}
\usepackage{orcidlink}
\usepackage{subfigure}
\usepackage{hyperref}
\hypersetup{colorlinks=true,
            linkcolor=blue,
            anchorcolor=blue,
            citecolor=blue}

\usepackage{chemfig}
\usepackage{bm}
\usepackage[version=3]{mhchem}

\begin{document}
\title{Study of nuclear magnetic resonance spectra with the multi-modal multi-level quantum complex exponential least squares algorithm}

\author{Antonio Márquez Romero
\orcidlink{0000-0002-1435-8017}
}
\email{antonio.marquezromero@fujitsu.com}
\affiliation{Fujitsu Research of Europe Ltd., Pozuelo de Alarcón, 28224 Madrid, Spain}

\author{Josh J.M. Kirsopp
\orcidlink{0000-0002-0048-1583}
}
\email{josh.kirsopp@fujitsu.com}
\affiliation{Fujitsu Research of Europe Ltd., Slough SL1 2BE, UK}

\author{Giuseppe Buonaiuto
\orcidlink{0009-0009-8427-292X}
}
\email{giuseppe.buonaiuto@fujitsu.com}
\affiliation{Fujitsu Research of Europe Ltd., Pozuelo de Alarcón, 28224 Madrid, Spain}

\author{Michal Krompiec
\orcidlink{0000-0002-2603-1070}
}
\email{michal.krompiec@fujitsu.com}
\affiliation{Fujitsu Research of Europe Ltd., Slough SL1 2BE, UK}

\begin{abstract}
    We present a novel application of the multi-modal, multi-level quantum complex exponential least squares (MM-QCELS) algorithm, a state-of-the-art, early fault-tolerant quantum phase estimation (QPE) technique, to the simulation and analysis of nuclear magnetic resonance (NMR) of spin systems. By leveraging the robustness and precision of MM-QCELS, we demonstrate enhanced phase resolution in quantum simulations of spin dynamics, also in systems with complex coupling topologies. Our approach enables accurate extraction of spectral features with up to an order of magnitude fewer evaluations of the time series signal in comparison with the conventional Fourier transform, making a significant step toward scalable quantum simulations of NMR Hamiltonians. This work bridges an advanced quantum algorithm design with a practical spectroscopic application, offering a promising new approach for a quantum-based chemical analysis.
\end{abstract}

\maketitle

\section{Introduction}
Nuclear Magnetic Resonance (NMR) is an essential spectroscopic technique widely used in chemistry and material science~\cite{levitt2008spin,keeler2011understanding}. It provides insightful information about the molecular structure, dynamics, and interactions between components. NMR is based on the probe of the local chemical environment of nuclei, most commonly single spins of hydrogen ($^1$H) or carbon ($^{13}$C), under the influence of a strong magnetic field, ideally achieving high spectral resolution. The response of the spin system under such electromagnetic pulses outputs the experimental spectra which we intend to compare with our theoretical descriptions. NMR enables us to determine the connectivity of atoms within a molecule, identify functional groups, and even infer three-dimensional structures. Beyond chemistry, NMR is also used in complex systems such as proteins~\cite{wuthrich1986nmr,wagner1993prospects,dyson2004unfolded}, where it can reveal dynamic processes, folding patterns, and intermolecular interactions. Furthermore, its quantitative outputs allows for precise concentration measurements in a mixed solution. 

Classical simulations of NMR are computationally challenging due to the exponential scaling of the spin Hamiltonian with the system size. Quantum computers are naturally suited for this kind of calculations, with a one-to-one correspondence between the number of spins in the system and the number of qubits in the quantum register. Notably, NMR analysis is considered one of the most promising candidates to demonstrate practical quantum advantage on this early stage of quantum hardware development~\cite{knill1998power,seetharam2023digital} and interest in the field has grown significantly in recent years~\cite{sels2020quantum,khedri2024impact,fratus2025can,ossorio2024simulating,burov2024towards,das2025exploring}.

In this work, we propose an application of a recently developed fault-tolerant method based on a single-ancilla quantum phase estimation routine for the simulation of spin-based NMR Hamiltonians. Similar to classical approaches, this application of the method relies on the evaluation of the time evolution of the system and computing the expectation value of spin operators. However, instead of an \textit{ad hoc} evaluation until the NMR spectra has the desired resolution, our application employs well-defined hyperparameters that provide an accurate estimate of the computational resources needed to resolve the NMR spectrum peaks within the requested resolution. We validate our implementation with molecular examples drawn from previous studies in the field, which also allow for classical-methods benchmarking, and we assess its viability in realistic settings. 

We do not claim that our implementation outperforms state‑of‑the‑art classical methods in practical runtimes. Rather, our contribution is to formulate an approach whose computational advantages emerge in the regime of fault‑tolerant quantum hardware. The value of our application is forward‑looking: once quantum processors are available and reliable, we expect the approach to enable tractable analysis in system sizes that are otherwise inaccessible to classical simulation, while also reducing experimental demands.

This paper is organized as follows: in Section~\ref{sec:NMR}, we introduce the fundamentals of NMR spectroscopy and discuss its treatment using classical methods. Section~\ref{sec:mmqcels} presents a summary of the original quantum algorithm and outlines its adaptation to the NMR spectroscopy problem. In Section~\ref{sec:quantum_circuits}, we detail the quantum circuit implementation of our proposed application. Section~\ref{sec:results} showcases the results of our simulations, and finally, in Section~\ref{sec:conclusions} we provide our concluding remarks and future prospects.

\section{The NMR spectroscopy} \label{sec:NMR}

Simulation of the NMR technique consists of the description of the local spin system of the molecule as the contribution of one-body and two-body terms using the Heisenberg-like Hamiltonian in one dimension\footnote{2D~\cite{simpson2011organic} and even 3D~\cite{spiess1991structure} studies of NMR spectroscopy can also be performed}
\begin{equation}\label{eq:NMR_ham}
\begin{split}
    H &= \sum_{i}^{N_{\rm{spins}}} \delta_i S_i^{(z)} + \sum_{i<j}^{N_{\rm{spins}}} J_{ij} (S_i^{(x)} S_j^{(x)} + S_i^{(y)} S_j^{(y)} + S_i^{(z)} S_j^{(z)}) \\ & = \frac{1}{2}\sum_{i}^{N_{\rm{spins}}} \delta_i Z_i + \frac{1}{4}\sum_{i<j}^{N_{\rm{spins}}} J_{ij} (X_i X_j + Y_i Y_j + Z_i Z_j),
\end{split}
\end{equation}
where $S_i^{(x,y,z)}$ are the spin operators acting on spin $i$, which are related to the usual Pauli operators $(X,Y,Z)$ as
\begin{equation}
        S^{(x)} = \frac{1}{2} X, \,\,\,\,
        S^{(y)} = \frac{1}{2} Y, \,\,\,\,
        S^{(z)} = \frac{1}{2} Z.
\end{equation}
The parameters of the NMR Hamiltonian are the chemical shifts $\delta_i$, which take into account an averaged effect of the surrounding cloud of the molecular electrons; and the coupling constants $J_{ij}$, which model the interaction between the different spins at play~\cite{smith1992hamiltoniansI, smith1992hamiltoniansII}. While other interaction terms may be included for a more accurate description, the isotropic Hamiltonian as described in Eq.~(\ref{eq:NMR_ham}) is commonly used in most liquid-state NMR experiments. For a system of $N_{\rm spins}$, there are a total of $N_{\rm spins}$ chemical shifts and $\frac{N_{\rm spins} (N_{\rm spins} - 1)}{2}$ coupling constants.

Under the action of a static magnetic field in the $Z$-axis direction, the individual spins experience a Zeeman effect and the chemical shifts are increased proportionally with the amplitude of such a magnetic field. The formulation of the NMR Hamiltonian in Eq.~(\ref{eq:NMR_ham}) is expressed in frequency units, but it is also a standard convention in the literature to write the one-spin term as $\gamma B S^{(z)}$, where $\gamma$ is the gyromagnetic ratio of the proton and $B$ is the strength of the magnetic field. For further details, we refer the reader to Ref~\cite{keeler2011understanding}.

\begin{figure}[t]
    \centering    \includegraphics[width=0.9\columnwidth]{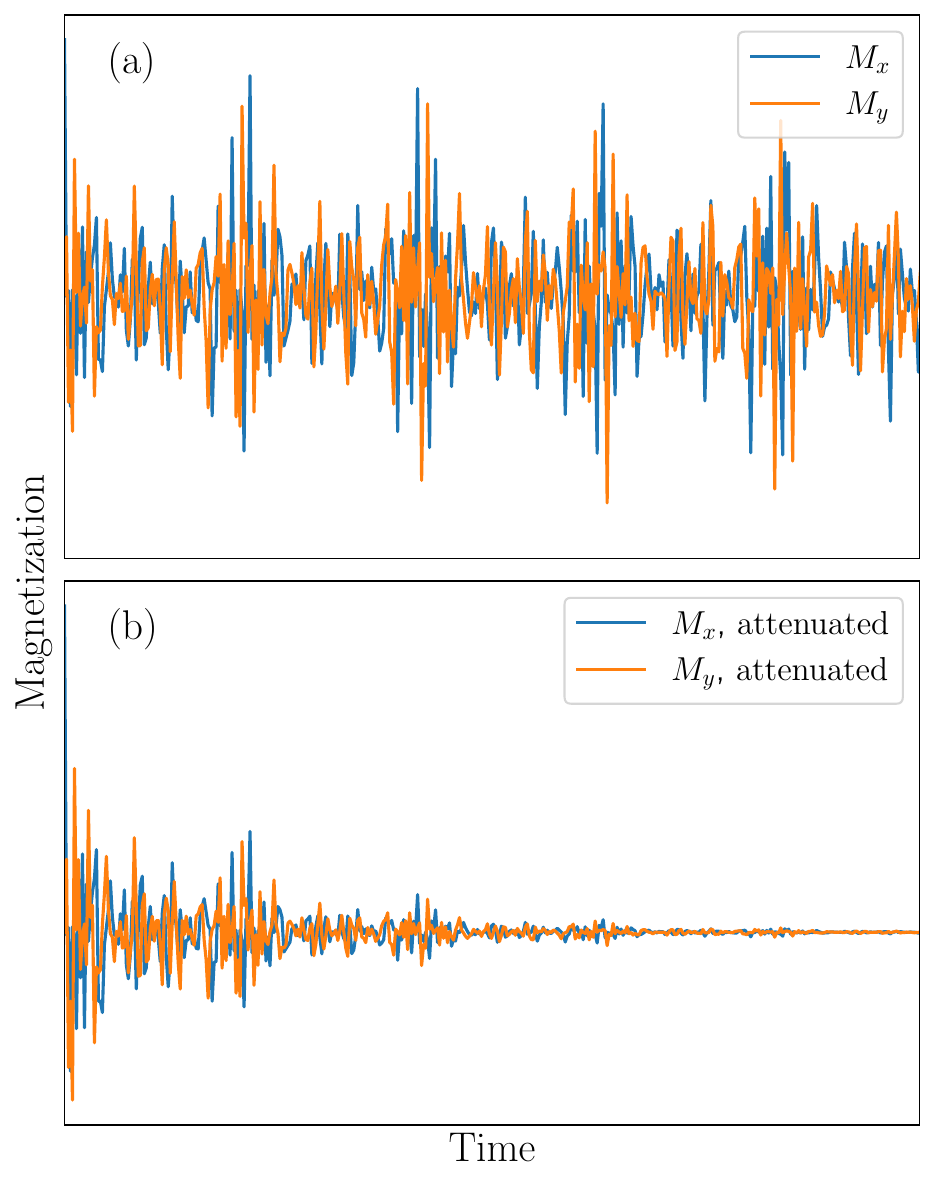}
    \caption{Standard plot of the magnetization signal. \textbf{a)} Case without attenuation, theoretical description of magnetization according to Eq.~(\ref{eq:M_eq}) \textbf{b)} Case with an attenuation factor of $\eta = 0.005\pi$ in units of inverse of time, included, corresponding to the FID signal in Eq.~(\ref{eq:FID}). Both subplots correspond to a simulation of a three-spin system, computed with $2^{14}$ temporal data points on a uniform grid with a time step of $\Delta t = 0.7$ in arbitrary time units. This resolution ensures accurate representation of the magnetization dynamics under the specified conditions.}
    \label{fig:magnetization}
\end{figure}

\begin{figure}[t]
    \centering    \includegraphics[width=0.9\columnwidth]{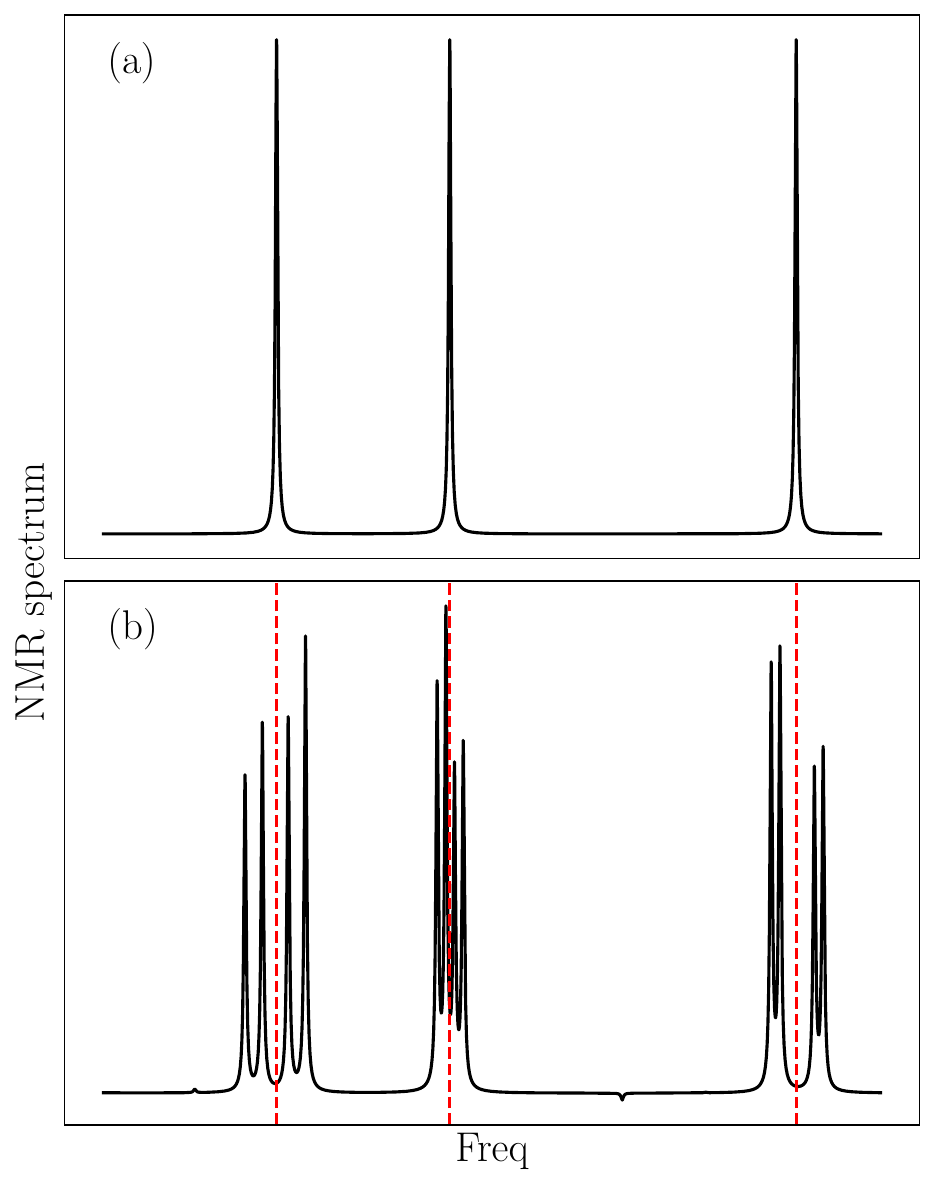}
    \caption{Standard plot of the NMR spectrum of a molecule with three spins. \textbf{(a)} Case of negligible coupling constants. Three peaks are observed centered at the values of the chemical shifts as described in the NMR Hamiltonian~(\ref{eq:NMR_ham}) for all $J_{ij} = 0$. \textbf{(b)} Case of non-negligible coupling constants. Three groups of four different peaks are centered around the three different chemical shifts. The peak splittings due to the two-spin interaction are equal to the difference in the coupling constants of the spins involved. Since in this case a spin can interact with the other two, there are two splittings observed within the same group.}
    \label{fig:nmr_standard}
\end{figure}

Once the sample is aligned along the $Z$-axis, a perpendicular electromagnetic pulse is applied, causing the different spins to precess around the $Z$-axis. Consequently, the collective motion creates a magnetization signal which can be measured. The magnetization $M$ on the perpendicular plane to the $Z$-axis can be computed as the expectation value of the $X-$ and $Y-$spin chains with respect to the state of the system $|\Psi\rangle$:
\begin{equation}\label{eq:M_eq}
\begin{split}
    \langle \Psi |M|\Psi\rangle &= \langle \Psi |M_x+iM_y|\Psi\rangle \\ &= \frac{1}{2}\sum_k^{N_{\rm{spins}}} \left(\langle \Psi |X_k|\Psi\rangle + i \langle \Psi |Y_k|\Psi\rangle\right). 
\end{split}
\end{equation}
Finally, if we consider a time evolution of the spin system given by the Hamiltonian~(\ref{eq:NMR_ham}) as 
\begin{equation}\label{eq:psit}
    |\Psi(t)\rangle = e^{-i H t / \hbar} |\Psi_0\rangle,
\end{equation}
where $|\Psi_0\rangle$ is the initial ansatz, the magnetization~(\ref{eq:M_eq}) is a time-dependent quantity and, for the sake of simplicity, we work in natural units where $\hbar = 1$. In the frequency domain, it corresponds to the NMR spectrum of the molecule, that is, 
\begin{equation}
\begin{split}
    \textrm{NMR}(f) &= \mathcal{F}\left[\langle \Psi(t) |M|\Psi(t)\rangle\right] \\ &= \mathcal{F}\left[\langle \Psi_0 |e^{i H t}Me^{-iHt}|\Psi_0\rangle\right]
\end{split}
\end{equation}
where $\mathcal{F}$ corresponds to the standard Fourier transformation over the positive domain
\begin{equation}
    \mathcal{F}[f(t)] = \int_0^{\infty} dt f(t) e^{2\pi i f t}.
\end{equation}
In a realistic experimental setting, the interaction of the sample with the environment needs to be taken into account. While a precise description needs the Lindblad formalism, phenomenologically we make the crude approximation of inserting a damping exponential factor in the magnetization signal, evolving in time as 
\begin{equation}\label{eq:FID}
    \langle \Psi(t) |M|\Psi(t)\rangle = e^{-\eta t}\langle \Psi_0 |e^{i H_{\rm{NMR}} t}Me^{-iH_{\rm{NMR}}t}|\Psi_0\rangle.
\end{equation}
We refer to this quantity as the Free Induction Decay (FID) signal~\cite{keeler2011understanding}. Figure~\ref{fig:magnetization} shows the standard magnetization signals in two subplots. Subplot $a)$ displays the signal without the damping factor, where the magnetization remains relatively stable over time. In contrast, subplot $b)$ illustrates the effect of the damping factor, causing the magnetization to decay exponentially until it approaches zero. 
Measured experimental NMR spectra have the form as shown in subplot $b)$ of Fig.~\ref{fig:nmr_standard}, where the peaks follow a Lorentzian lineshape, that is, 
\begin{equation}\label{eq:lorentzian}
    L(f) = A\frac{\eta}{\eta^2 - (f - f_0)^2},
\end{equation}
where $A$ and $f_0$ are the amplitude (height) and the center (position of the height) of the peak, respectively. Thus, the peak amplitudes and centers are obtained by means of lineshape analysis~\cite{roy2002esprit,hua2002matrix}. The centers are directly related to the chemical shifts and coupling constants in the NMR Hamiltonian~(\ref{eq:NMR_ham}), as we show in the Supplementary Information. In this work, for the initial state ansatz we always suppose the sample is prepared in the pure Hadamard state $|+\rangle^{\otimes N_{\rm{spins}}} = \left[ \frac{1}{\sqrt{2}} (|0\rangle + |1\rangle) \right] ^{\otimes N_{\rm{spins}}}$, which is consistent with previous works in this field~\cite{burov2024towards}, in contrast to a density-operator formalism to obtain the expectation values of the magnetization operator~\cite{khedri2024impact,fratus2025can,ossorio2024simulating}.

The Hadamard state was chosen as the initial ansatz because it creates a uniform superposition over all computational basis states, providing an unbiased starting point for the algorithm. In addition, Hadamard gates are among the simplest single-qubit operations and are part of the standard universal gate set, making them straightforward to implement on both quantum hardware and simulators.

While standard quantum algorithms such as the Variational Quantum Eigensolver (VQE)~\cite{fedorov2022vqe,tilly2022variational} and Quantum Phase Estimation (QPE) have been applied to NMR analysis~\cite{ossorio2024simulating}, in this work we adopt a different approach: we compute expectation values of the magnetization time‑series in Eq.~(\ref{eq:FID}) and extract the relevant spectral properties through classical post‑processing using the MM‑QCELS framework described in Section~\ref{sec:mmqcels}. 

\section{The MM-QCELS algorithm}\label{sec:mmqcels}

The multi-modal multi-level quantum complex exponentials least squares (MM-QCELS)~\cite{ding2023simultaneous} is an optimization method originally developed to extract multiple eigenvalues of a given Hamiltonian. It belongs within the single-ancilla quantum phase estimation myriad of methods~\cite{kohne2024single,nelson2024assessment}. In this Section, we include the key aspects of the method required for completeness, and direct the interested reader to the original work by Ding et al.~\cite{ding2023simultaneous} for a full technical description. It is based on the Hamiltonian evolution of an ansatz written in the eigenbasis
\begin{equation}
    e^{-i H t} |\Psi\rangle =  \sum_i c_i e^{-i H t} |E_i\rangle = \sum_i c_i e^{-i E_i t} |E_i\rangle, 
\end{equation}
where $|E_i\rangle$ is the eigenstate with energy $E_i$ and $c_i$ are the coefficients of the eigendecomposition. The chosen ansatz $|\Psi\rangle$ needs to be expressive enough to contain all eigenstates of interest. 

The algorithm proceeds by generating a dataset $[t_i, O_i=O(t_i)]$, computing expectation values of the time evolution operator
\begin{equation}\label{eq:expval_evol}
    O(t) = \langle \Psi|e^{-iHt}|\Psi\rangle
\end{equation}
for times $t_i$ drawn randomly from a truncated Gaussian probability distribution of the form
\begin{equation}\label{eq:trunc_gauss}
    G_{T}(t) = 
    \begin{cases}
        \frac{1}{N_{G_T} T} e^{-\frac{t^2}{2T^2}},  & t \in [-T,T] \\
        0, & \textrm{otherwise}
    \end{cases}
\end{equation}
where $N_{G_T}$ is a normalization factor. Using this dataset, an optimization problem is built with the following cost function
\begin{equation}\label{eq:original_cost_function_MMQCELS}
    \mathcal{L}(\bm{r}, \bm{\theta}) = \frac{1}{N}\sum_{n=1}^{N} \left|\langle \Psi|e^{-iHt_n}|\Psi\rangle  - \sum_k r_k e^{-i \theta_k t_n}\right|^2.
\end{equation}
Employing these ingredients, the algorithm consists in an iterative procedure as follows:
\begin{enumerate}
    \item Generate $N_0$ time points drawn randomly from a truncated Gaussian distribution with time $T_0$ and build the expectation value of the Hamiltonian operator, Eq.~(\ref{eq:expval_evol}), dataset. 
    \item Minimize the cost function in Eq.~(\ref{eq:original_cost_function_MMQCELS}) using the initial frequency bounds (-$\pi$, $\pi$) for all $\bm{\theta}$.
    \item Using the optimal solution $(\bm{r}_{opt}, \bm{\theta}_{opt})$ from the minimization, a new iteration $j$ is performed. Now the bounds for the parameters are updated by  $(\bm{\theta}_{opt} - \frac{\pi}{T_j}, \bm{\theta}_{opt} + \frac{\pi}{T_j})$ and the initial guess is the optimal solution found before. The dataset is generated with new randomly sampled times with a truncated Gaussian distribution~(\ref{eq:trunc_gauss}) with parameters $T_j = 2^jT_0$.
    \item The procedure is repeated iteratively until convergence and/or the number of estimated maximum iterations by the algorithm is reached. 
\end{enumerate}
Since the bounds decrease exponentially with the number of iterations, only a few iterations are necessary to obtain all relevant eigenvalues, given an expressive enough ansatz $|\Psi\rangle$. The number of samples $N$, initial time $T_0$ and an estimate for the number of iterations $J$, are, respectively, given by:
\begin{equation}\label{eq:hyperparameters}
\begin{split}
    N & = \frac{1}{\sqrt{\varepsilon}},\\
    T_{0}&=2^{-[1 + \log_{2}(1/\epsilon)] \delta / (N\epsilon)}, \\
    J&=\log_{2}(1/\varepsilon) + 1,
\end{split}
\end{equation}
where $\varepsilon$ is the target error in the estimation of the peaks, $\delta=\Theta(\sqrt{1-p})$, being $p$ the overlap between the target quantum state to be used for the estimation of the expectation values and the ansatz deployed in the algorithm, and $\Theta(\cdot)$ is the step function. In this work, we assume that $\delta=1$ and $p=0.1$. We observe that both the number of samples and iterations needed are primarily a function of the desired accuracy. This offers a crucial advantage when increasing the size of the system, as these numbers do not inherently scale with it. While a more complex system might require a higher desired accuracy to resolve all features, the algorithm itself remains efficient.
The minimization is performed with the L-BFGS-B method~\cite{byrd1995limited,zhu1997algorithm}. The quantum part of the algorithm is used exclusively to obtain expectation values of the target operators, as in Eq.~(\ref{eq:expval_evol}), providing the spectral information that the classical post‑processing then reconstructs. We provide further details of this quantum component and the construction of quantum circuits for the NMR application in Section~\ref{sec:quantum_circuits}.

\subsection{Adjusting MM-QCELS for NMR spectral analysis}

The MM-QCELS algorithm described in the former section can be modified to handle NMR datasets from the magnetization time-series signals. We use a cost function of the form 
\begin{equation}\label{eq:cost_function_MMQCELS}
\begin{split}
    \mathcal{L}(\bm{r}, \bm{\theta})  = \frac{1}{N}\sum_{n=1}^{N} & \bigg|\langle \Psi(t_n)|(M_x+iM_y)|\Psi(t_n)\rangle -  \\ & - \sum_k r_k e^{-i \theta_k t_n}\bigg|^2,
\end{split}
\end{equation}
where $|\Psi(t)\rangle$ is a time-evolved wavefunction as in Eq.~(\ref{eq:psit}). The optimal parameters $(\bm{r}_{opt},\bm{\theta}_{opt})$ found by MM-QCELS have the physical significance as the amplitudes and positions of the peaks, respectively. The only difference in the full pipeline of the algorithm, with respect to the original method, is to constrain the bounds so that the parameters are always positive. A major drawback of this implementation is the exponentially-scaling number of peaks found in the NMR spectrum, which goes as $N_{\rm{spins}} 2^{N_{\rm{spins}}-1}$ (assuming all peaks can be resolved). This problem is also shared in the original implementation, as the number of eigenvalues in a Hamiltonian also increases exponentially with the number of qubits needed to encode it. A partial solution is to introduce a limiting number $K$ of peaks to be found in the exponential expansion of the cost function~(\ref{eq:cost_function_MMQCELS}), or to constrain the time domain to find only desired, specific peaks. A promising alternative, independent of $K$, is based on the Quantum Multiple Eigenvalue Gaussian Search (QMEGS) method~\cite{Ding:2024qvu}, and it will be the subject of future work. 

\section{Circuit preparation}
\label{sec:quantum_circuits}

\begin{figure}
    \centering
    \includegraphics[width=0.9\columnwidth]{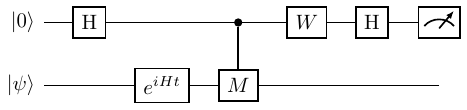}
    \caption{The general form of the Hadamard test circuits used in this publication to generate the dataset, with a controlled-magnetization gate using a single ancilla qubit. $W$ is the identity gate or the $S^{\dagger}$ gate (with $S$ being the standard phase gate) if we want to measure the real or imaginary part of the magnetization operator, corresponding to the $M_x$ and $M_y$ components, respectively.}
    \label{fig:hadamardtest}
\end{figure}

\begin{figure}
    \centering
    \includegraphics[width=0.9\columnwidth]{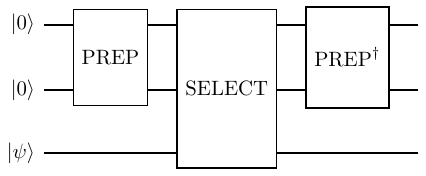}
    \caption{The general form of the block encoding circuits used in this publication to encode the magnetization operator. Given the operator to be encoded as a linear combination of unitaries, the PREP operator encodes the coefficients and the SELECT operator encodes the unitaries.}
    \label{fig:be_circuit}
\end{figure}

\begin{figure}
    \centering
    \includegraphics[width=0.9\columnwidth]{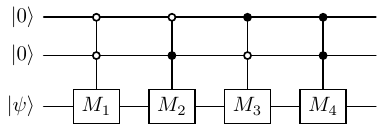}
    \caption{The general form of the SELECT operator circuits used in this publication to encode the magnetization operator. The different controlled states specify which term $M_i$ of the operator to be applied to $|{\psi}\rangle$}
    \label{fig:select_circuit}
\end{figure}

Central to the MM-QCELS algorithm is the evaluation of a dataset comprising a series of expectation values of the time evolution operator at different times via Hadamard test circuits. The main modification we make to the workflow from the original publication is that we evaluate the expectation value of the magnetization operator with respect to a time-evolved state instead. The circuits therefore consist of two main blocks, as shown in Fig~\ref{fig:hadamardtest}. 
In order to adequately capture the dynamics of the system, we approximate the time-evolution operator, $U=e^{iHt}$, using an $6^\text{th}$-order Trotter product formula of the form introduced by  Yoshida~\cite{yoshida1990construction} and further developed by Morales et al.~\cite{Morales2025Trotter}, which we hereafter refer to as Yoshida-Trotter formula. Moreover, since we are not attempting to solve for the eigenvalues of the magnetization operator, and this operator is a sum of spin operator chains, we use the block-encoding approach of a linear combination of unitaries as reported by Childs and Wiebe~\cite{childs2012hamiltonian} for its encoding in the circuit. The block encoding circuit consists of PREP and SELECT operators, as shown in Fig.~\ref{fig:be_circuit}, whose action on a quantum state is
\begin{equation}\label{eq:PREP}
    \textrm{PREP}|0\rangle^{n_a} = \sum_i \sqrt{\frac{|{a_i}|}{\lambda}} |i\rangle,
\end{equation}
\begin{equation}
    \textrm{SELECT}|i\rangle\otimes|\psi\rangle = |i\rangle\otimes M_i|\psi\rangle,
\end{equation}
where $n_a$ is the number of ancilla qubits needed for the block encoding and $a_i$ are the coefficients corresponding to the different terms in the operator to be encoded, with $\lambda=\sum_i |a_i|$ being the 1-norm of the coefficients. The $n_a$ number is equal to $\left\lceil \log_2(N_{\rm{terms}})\right\rceil$ where $N_{\rm{terms}}$ is the number of terms in the operator to be encoded, which in the case of the magnetization operator, equals twice the number of spins.
The PREP operator~\cite{shende2005synthesis} encodes the coefficients $a_i$ of the magnetization operator in Eq.~(\ref{eq:M_eq}). For this case, all the coefficients $a_i$ are equal to $\frac{1}{2}$ and Eq.~(\ref{eq:PREP}) simplifies to
\begin{equation}
    \textrm{PREP}|0\rangle = \frac{1}{\sqrt{2N_{\textrm{spins}}}}\sum_i |i\rangle.
\end{equation}
The SELECT operator applies the specific term $M_i$ in the magnetization operator corresponding to the coefficient $a_i$.  
The SELECT operator acts on the quantum state by activating respective unitaries of the operator by means of multi-controlled gates as shown in Fig.~\ref{fig:select_circuit} for the case of four unitaries.

\section{Results and discussion}
\label{sec:results}

A crucial step in the MM-QCELS algorithm is obtaining the magnetization signal from the sampled output of Hadamard test circuits using a time-evolved ansatz in the quantum simulations of the spin chains. The Trotterized approximation for the time evolution operator is hence necessary for the construction of the signal: in this work, we first run extensive numerical experiments to estimate the error we should expect for different Trotter orders and number of steps, as a function of the signal length. It is sensible to expect that the larger the signal, the longer times in the evolution will be sampled out with the circuit simulations, and thus the larger the expected error between the signal generate with the exact time evolution operator in place and the one obtained via Trotter:
\begin{equation}
    \label{eq:trotter_error}
    \mathcal{R}=||S-S_\text{Trotter}||_{2}.
\end{equation}
In Eq.~(\ref{eq:trotter_error}) $||\cdot||_{2}$, is the 2-norm between vectors representing the signal obtained with the exact unitary evolution operator, $S$, and the one obtained via Trotter, $S_\text{Trotter}$. We analyze this quantity considering four different Trotter expansions, namely the first, second and fourth order Trotter-Suzuki, and the sixth order Yoshida-Trotter~\cite{yoshida1990construction,Morales2025Trotter}. The latter is computationally less expensive than the analogous expansion made with the standard Trotter-Suzuki. We further consider three different Trotter steps, i.e., the number of repeated application of the expanded operator, to assess the effect on the reconstruction of the signal. 
\begin{figure*}[ht]
    \centering
    \includegraphics[width=0.9\textwidth]{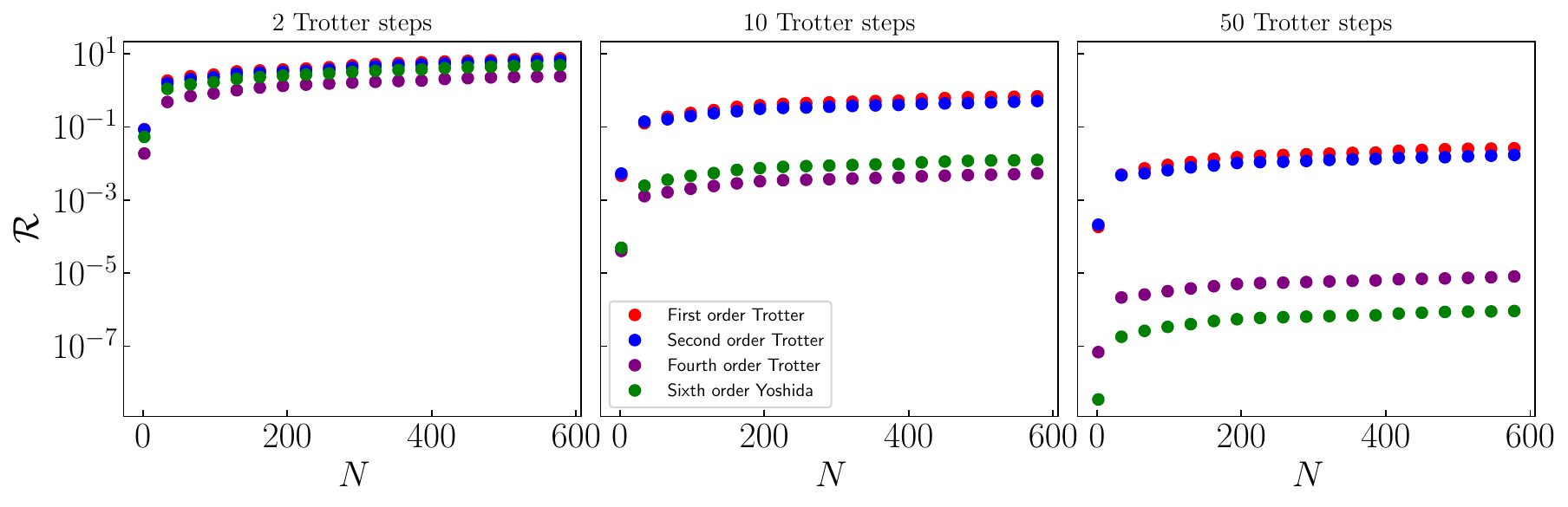}
    \caption{Error of the approximated magnetization signal dataset, obtained with Trotter expansion of the evolution operator, with respect to the exact case and as a function of the number of samples $N$. Data are generated from a two-spins Hamiltonian with coefficients $(\delta_0, \delta_1, J_{01}) = (4 \,\, \textrm{ppm},6 \,\, \textrm{ppm},2 \,\, \textrm{Hz})$. Samples are generated fixing $T_j=20\,\, s$ in the MM-QCELS pipeline. }
    \label{fig:trotter_error}
\end{figure*}
Results are shown in Fig.~\ref{fig:trotter_error}: we can observe that, with respect to an increasing of the signal length $N$, there is a general trend of degradation of the reconstruction, although it saturates quickly. Two Trotter steps are insufficient to reach a good quality of the signal, furthermore, in this regime, the Trotter order is almost not affecting the approximation. On the contrary, the other two regimes considered, using $10$ and $50$ steps, are quite sensitive to the Trotter order, showing an exponential suppression of the error $\mathcal{R}$ for Trotter orders greater than $2$.

These results form a set of prescriptions, when designing the experimental pipeline for the estimation of peaks. Once the target signal length, $N$, is estimated, a selection of the optimal Trotter order and steps can be performed, aiming at minimizing the computational cost of the circuit, without losing an excessive amount of information in the reconstruction of the signal.

\subsection{Sulfanol}
\begin{figure}
    \centering
    \includegraphics[width=0.9\columnwidth]{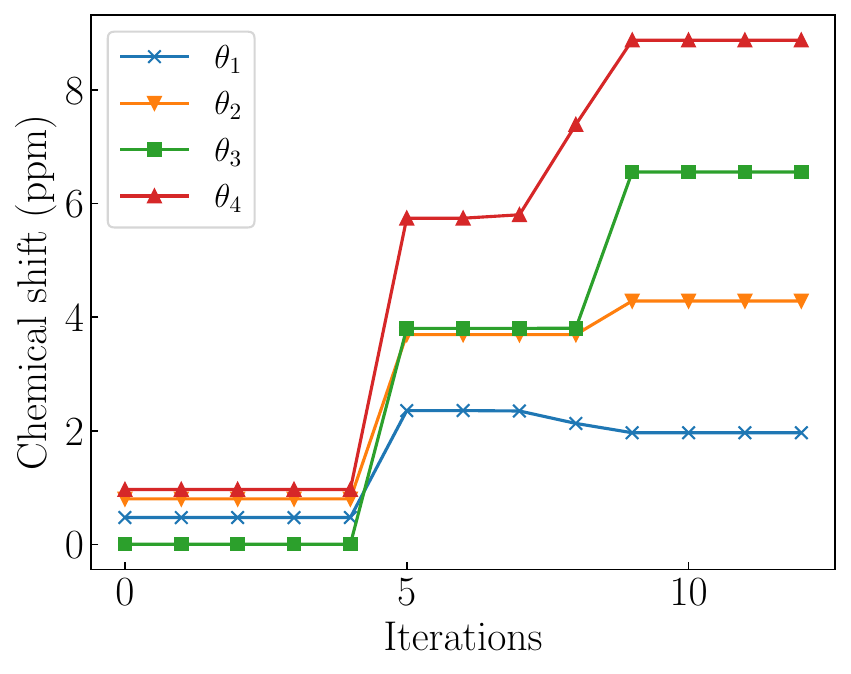}
    \caption{Convergence of the optimal peak positions $\theta_k$ in the cost function~(\ref{eq:cost_function_MMQCELS}) with the iterative MM-QCELS pipeline for sulfanol.}
    \label{fig:sulfanol_nmr}
\end{figure}

Sulfanol is a molecule with two hydrogen atoms with formula \ce{HSOH}. 
\begin{figure}[hb]
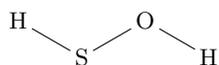

    \centering
\chemfig{*6(H-S-O*6(-H))}
    \caption{Sulfanol molecule.}
    \label{fig:sulfanol_molecule}
\end{figure}
To simulate its $^1$H NMR spectrum, we model it as a two-spin system using the following chemical shifts and coupling constant parameters~\cite{ossorio2024simulating}
\begin{equation}\label{eq:sulfanol_pars}
\begin{split}
    \delta_0 & = 3.44 \, \rm{ppm},\\
    \delta_1 & = 7.40 \, \rm{ppm}, \\
    J_{01} & = 2.32 \, \rm{Hz},
\end{split}
\end{equation}
and we assume a spectrometer applying a very low field of roughly 0.02 T, corresponding to a reference resonance frequency of 1 $\rm{MHz}$\footnote{The resonance frequency $\nu$ and the magnetic field $B$ are related by the gyromagnetic ratio $\gamma$ as $\nu = \gamma B$. The gyromagnetic ratio for protons is $\gamma = 42.577\,\, \rm{MHz/T}$~\cite{levitt2008spin}. }. Because the coupling constant is not large in comparison with the difference in chemical shifts, a large magnetic field in order to increase the ratio between them is not needed. We would like to emphasize that operating at low magnetic fields is one of the main advantages of our method. High-field spectrometers require expensive superconducting magnets to achieve large magnetic fields, whereas our approach significantly reduces this requirement, leading to substantial savings and simpler infrastructure. A rescaling of the whole Hamiltonian for the MM-QCELS was not applied in this case either. We chose for the simulation a chemical shift resolution of $\varepsilon = 0.001 \,\, \textrm{ppm}$, which results in hyperparameters~(\ref{eq:hyperparameters}) of $J=12$ iterations, $N=32$ datapoints and $T_0=0.0015 s$ for the initial time step. We run the MM-QCELS pipeline and store the optimal solutions given by the minimizer at every iteration. We plot the resulting phases, corresponding to the peak centers, in Fig.~\ref{fig:sulfanol_nmr} and the fully-converged results are summarized in Table~\ref{tab:sulfanol}.

\begin{figure}
    \centering
    \includegraphics[width=0.9\columnwidth]{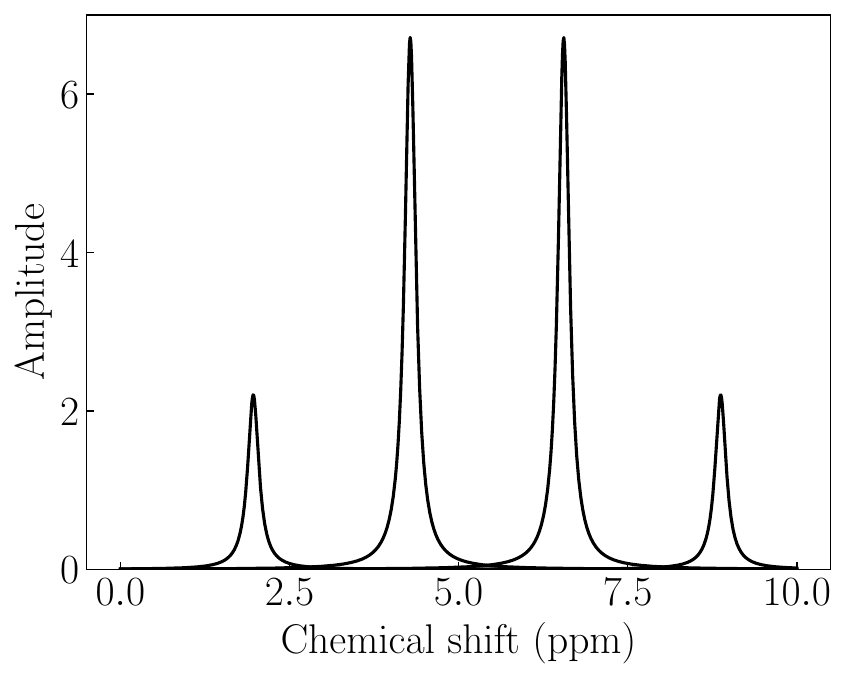}
    \caption{$^1$H NMR spectrum for sulfanol using the parameters from Table~\ref{tab:sulfanol} and using an attenuation factor of $\eta=0.1\,\,s^{-1}$ in the Lorentzian line function~(\ref{eq:lorentzian}).}
    \label{fig:sulfanol_spectrum}
\end{figure}

\begin{table}
\centering
\caption{Final converged results for the amplitudes and centers of the peaks obtained for sulfanol after the full MM-QCELS simulation. Amplitudes are normalized to 1.}
\label{tab:sulfanol}
\begin{tabular}{|c|c|c|c|}
\hline
$\#$ & Amplitude & Center (ppm) & Diff (Hz) \\
\hline
$p_1$ & 0.22067 & 1.96522 & \multirow{2}{*}{2.32000} \\
\cline{1-3}
$p_2$ & 0.67179 & 4.28522 & \\
\hline
$p_3$ & 0.67179 & 6.55477 & \multirow{2}{*}{2.32000} \\
\cline{1-3}
$p_4$ & 0.22067 & 8.87477 & \\
\hline
\end{tabular}
\end{table}
We observe a remarkable precision in the difference of the observed peak splittings, which correspond to the coupling constant $J_{01}$ in Eq.~(\ref{eq:sulfanol_pars}). This level of agreement, together with the smooth convergence seen in Fig.~\ref{fig:sulfanol_nmr}, indicates that the extracted peak positions and amplitudes correspond to those expected in an experimental $^1$H NMR spectrum of sulfanol. Using standard Lorentzian functions~(\ref{eq:lorentzian}), we can plot this spectrum with the converged results, as shown in Fig.~\ref{fig:sulfanol_spectrum}. 
Finally, one of the most compelling aspects of this approach is its ability to produce accurate spectral estimates using significantly fewer data points than conventional methods. Classical techniques involving the Fourier transform often require dense sampling and long acquisition times which translates into roughly $2^{12-16}$ data point evaluations. MM-QCELS achieves comparable or superior resolution with up to an order of magnitude fewer evaluations of the magnetization. For the case of sulfanol, roughly $J \cdot N = 12 \cdot 32 = 384$ data points are necessary. This efficiency not only reduces computational overhead but also makes the method particularly well-suited for quantum simulations, where data acquisition is inherently costly. 

\subsection{Cis-3-chloroacrylic acid}

\begin{figure}
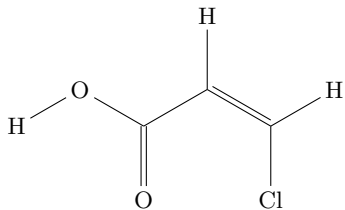

\centering
\chemfig{*6( O(-H)-(=O)-*6(=(-Cl)-H  ) -H)}
    \caption{Cis-3-chloroacrylic acid.}
    \label{fig:c3h3o2cl_molecule}
\end{figure}

\begin{figure}
    \centering
    \includegraphics[width=0.9\columnwidth]{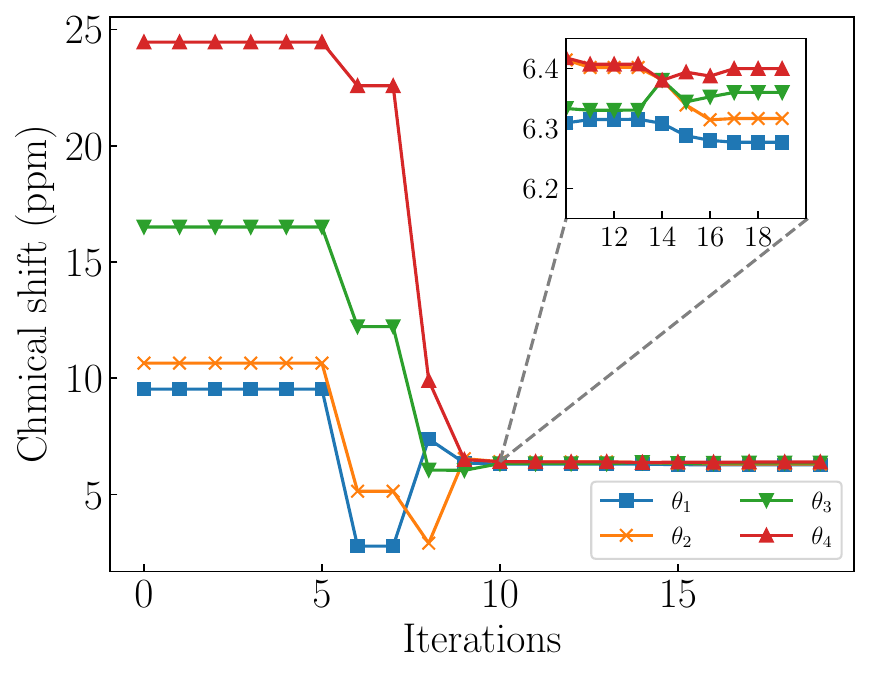}
    \caption{Convergence of the optimal peak positions $\theta$ for cis-3-chloroacrylic acid in the MM-QCELS algorithm. An inset zoom plot has been placed to better observe the convergence of the results.}
    \label{fig:cis3ch_nmr}
\end{figure}

The cis-3-chloroacrylic is a molecule with three hydrogen atoms with the formula \ce{C3H3O2Cl} and structure shown on Fig~\ref{fig:c3h3o2cl_molecule}. To obtain its $^1$H NMR spectrum, we model it as a two-spin system\footnote{Although the molecule has three inequivalent protons, the acidic proton is not coupling with the other two and may not be visible on the spectrum due to rapid exchange with the solvent} using the following chemical shifts and coupling constant parameters~\cite{khedri2024impact} 
\begin{equation}\label{eq:cis_pars}
\begin{split}
    \delta_0 & = 6.375 \, \rm{ppm},\\
    \delta_1 & = 6.302 \, \rm{ppm}, \\
    J_{01} & = 7.92 \, \rm{Hz}.
\end{split}
\end{equation}
We can observe in this case the higher resolution needed in order to be able to identify the individual peaks as the chemical shifts are very close together and the coupling constant is large in comparison. The MM-QCELS pipeline is adjusted by requesting a higher resolution in the phases of the cost function in Eq.~(\ref{eq:cost_function_MMQCELS}). In addition, for a smoother optimization we also assume a spectrometer applying a field of 4.7 T corresponding to a resonance frequency of 200 $\rm{MHz}$, and a global rescaling of the Hamiltonian dividing by a constant factor of 2000. Again, the intensity of the magnetic field is much lower than what is reported in this case for a similar accuracy~\cite{khedri2024impact}. The chemical shift resolution we chose for this case is $\varepsilon=10^{-5} \,\, \textrm{ppm}$, which in turn yields the hyperparameters~(\ref{eq:hyperparameters}) of $J=19$ iterations, $N=316$ datapoints and $T_0=0.00015 \,\, s$. Similarly as with sulfanol, we store the optimal results for the peak centers at every iteration and we plot them in Fig.~\ref{fig:cis3ch_nmr}.  
The fully-converged results are summarized in Table~\ref{tab:cis}
\begin{table}
\centering
\caption{Final converged results for the amplitudes and centers of the peaks obtained for cis-3-chloroacrylic acid after the full MM-QCELS simulation. Amplitudes are normalized to 1.}
\label{tab:cis}
\begin{tabular}{|c|c|c|c|}
\hline
$\#$ & Amplitude & Center (ppm) & Diff (Hz) \\
\hline
$p_1$ & 0.23712 & 6.27717 & \multirow{2}{*}{7.92} \\
\cline{1-3}
$p_2$ & 0.66821 & 6.31677 & \\
\hline
$p_3$ & 0.66468 & 6.36022 & \multirow{2}{*}{7.92} \\
\cline{1-3}
$p_4$ & 0.23553 & 6.39982 & \\
\hline
\end{tabular}
\end{table}

\begin{figure}
    \centering
    \includegraphics[width=0.9\columnwidth]{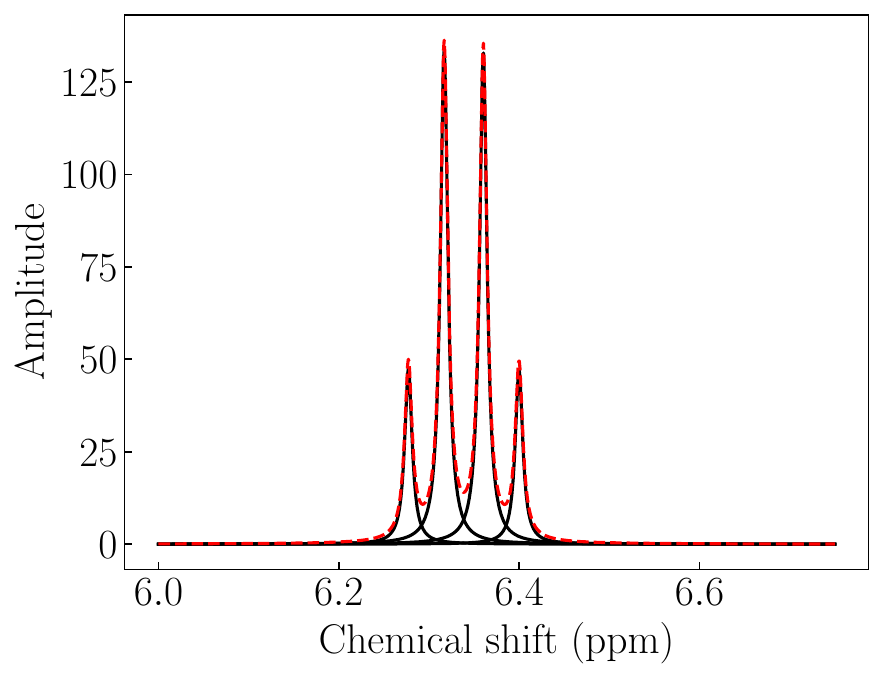}
    \caption{$^1$H NMR spectrum for cis-3-chloroacrylic acid using the parameters from Table~\ref{tab:cis} and with an attenuation factor of $\eta = 0.005 \,\, s^{-1}$ in the Lorentzian line function~(\ref{eq:lorentzian}). The sum of lineshape profiles, which would correspond to a realistically observed NMR spectrum, is also plotted using dashed lines.}
    \label{fig:cis_spectrum}
\end{figure}

Even in this case when a large resolution is needed, MM-QCELS performs significantly well and is able to extract the spectral features of the observed peaks and show a reliable convergence. The difference in the peaks need to be rescaled back by the spectrometer field to obtain the predicted coupling constant, 
\begin{equation}
    J_{01}^{\rm{pred}} = 200\,\, \rm{MHz} \,\,\cdot \,\, 0.0396 \,\, \rm{ppm} = 7.92 \,\, \rm{Hz}, 
\end{equation}
which is exactly the coupling constant used in the simulation~(\ref{eq:cis_pars}). Again, using Lorentzian functions~(\ref{eq:lorentzian}) with the converged results in Table~\ref{tab:cis}, we can plot the $^1$H NMR spectrum as shown in Fig.~\ref{fig:cis_spectrum}.
We notice in the $^1$H NMR spectrum of cis-3-chloroacrylic acid of Fig.~\ref{fig:cis_spectrum} the very small attenuation factor used in order to discern the peaks. While this factor needs to be included in a damping exponential to the magnetization to make the FID~(\ref{eq:FID}) in classical approaches using the Fourier transform, in contrast MM-QCELS can operate using the raw magnetization in the cost function~(\ref{eq:cost_function_MMQCELS}), without any limitations.

While we were mostly focused on the peak centers, which indeed provide the most substantial information as the chemical shifts and coupling constants can be estimated from there, MM-QCELS also includes the peak amplitudes in the full pipeline optimization. In the plots of the $^1$H NMR spectra of sulfanol in Fig.~\ref{fig:sulfanol_spectrum} and cis-3-chloroacrylic acid in Fig.~\ref{fig:cis_spectrum} we can observe the so-called roofing effect. This effect refers to an asymmetric intensity distribution observed in coupled spin systems after the peak splitting caused by the interaction. In a two-spin system (like in this study), the split doublets can show unequal peak heights and the inner peaks (the ones that are closer together) are often taller than the outer ones. This creates a roof-like appearance in the spectrum, hence the name. MM-QCELS is also able to capture that phenomenon.

\section{Conclusions}
\label{sec:conclusions}

In this work, we have demonstrated the successful application of the MM-QCELS algorithm to the analysis of $^1$H NMR spectra, achieving high-resolution identification of peaks even in scenarios involving closely spaced chemical shifts. The method consistently outperforms traditional Fourier transform approaches in terms of resource efficiency. MM-QCELS proved capable of extracting spectral features with minimal quantum resources, making it a compelling candidate for scalable, quantum-enhanced, spectroscopy. We find that significantly lower magnetic field strengths are needed compared to conventional high-field NMR techniques. This reduction could have profound practical implications as high-field superconducting magnets represent one of the most expensive components of modern spectrometers. By enabling accurate spectral analysis at lower fields, our approach could make advanced NMR capabilities more accessible to a broader range of laboratories and applications.

A key limitation identified in this work is the classical optimization step used to identify all possible spectral peaks. This step becomes increasingly computationally expensive as the number of spins in the system grows, exhibiting exponential scaling. A workaround to this scalability issue is to limit the optimization to a predefined number of peaks within a specific frequency window. Rather than resolving the entire spectrum, this approach focuses on regions of interest. This strategy is commonly employed in practice, as prior information about the sample or molecular composition is often available and can guide the selection of relevant spectral regions. Addressing this challenge will be crucial for extending the applicability of MM-QCELS. Broadly, compressed sensing~\cite{tang2013compressed,duarte2013spectral} and quantum filtering~\cite{sakuma2025quantum,lu2021algorithms,zeng2021universal,sakuma2024entanglement} techniques may offer a promising solution for reducing the dimensionality of spectral reconstruction problems, especially when the signal is sparse in the frequency domain, by enabling targeted peak identification. In addition, Green’s functions~\cite{endo2020calculation,libbi2022effective,dhawan2024quantum,piccinelli2025efficient,aychet2025quantum} offer a powerful framework for NMR analysis by capturing the system’s spectral response. Nevertheless, the robustness of MM-QCELS reinforces the value of fault-tolerant quantum phase estimation techniques in practical quantum chemistry and spectroscopy applications. The range of applications of the approach presented in this work, fall beyond NMR as, in practice, the use of MM-QCELS allows the construction of the signal with fewer data, in the post-processing pipeline, than other approaches: hence this work foster the possibility of reducing the size of data needed for training quantum or quantum-inspired machine learning algorithm, especially those that require the evaluation of a large amount of expectation value to function properly, such as Quantum Kernels and Quantum Neural Networks ~\cite{springerQuantumKernel,natureTrainingDeep}.  

In future work, we will explore the integration of the QMEGS algorithm, which may offer complementary advantages as it partly bypasses the classical optimization involved in MM-QCELS, lowering the overall computational cost of the algorithm. Additionally, we plan to investigate the use of qubitization-based Hamiltonian simulation~\cite{qubitization} as an alternative to the Trotter-Suzuki decomposition employed in this study. This shift could further reduce circuit depth and improve scalability, paving the way for more efficient quantum simulations of NMR phenomena.

\section*{Author contributions}

Conceptualization: AMR; methodology: AMR, JK, GB, MK; data curation: AMR, JK, GB; investigation: AMR; formal analysis: AMR; validation: AMR, JK, GB; writing – original
draft: AMR; writing – review \& editing: AMR, JK, GB, MK.

\section*{Conflicts of interest}
There are no conflicts to declare.

\section*{Data availability}

All data supporting the findings of this study were obtained from the references cited throughout the manuscript. The implementation details of the algorithm are fully described in the text, ensuring reproducibility and transparency.

\section*{Acknowledgements}

We would like to express our gratitude to our colleagues at Fujitsu Research of Japan for the time and effort they dedicated to reviewing this manuscript.

\appendix

\section{NMR spectrum under a high magnetic field regime}

When the amplitude of the applied magnetic field is large enough, the chemical shifts in the NMR Hamiltonian are considerably larger than the coupling constants between the $^1$H spins. Thus, they can be disregarded and write the NMR Hamiltonian under this regime as 
\begin{equation}
    H = \frac{1}{2}\sum_i \delta_i Z_i.
\end{equation}
Using this Hamiltonian, the time-evolved magnetization signal, and consequently the NMR spectrum, can be exactly computed,
\begin{equation}
\begin{split}
    \langle \Psi(t)|M|\Psi(t)\rangle &= \langle \Psi_0|e^{iHt}M_xe^{-iHt}|\Psi_0\rangle \\ & + i\langle \Psi_0|e^{iHt}M_ye^{-iHt}|\Psi_0\rangle,
\end{split}
\end{equation}
and the rotation of Pauli matrices has a compact expression, exploiting its product properties, 
\begin{equation}
\begin{split}
    e^{i \theta Z} X e^{-i \theta Z} &= \cos(2\theta) X - \sin(2\theta) Y, \\
    e^{i \theta Z} Y e^{-i \theta Z} &=  \cos(2\theta) Y + \sin(2\theta) X, \\  
    e^{i \theta Z} Z e^{-i \theta Z} &= Z.
\end{split}
\end{equation}
Thus, we have 
\begin{equation}
\begin{split}
    \langle \Psi(t)|M|\Psi(t)\rangle &= \sum_k \big[ \cos(\delta_k t) \langle X_k\rangle - \sin(\delta_k t) \langle Y_k \rangle  \\&+ i \sin(\delta_k t) \langle X_k \rangle + i \cos(\delta_k t) \langle Y_k\rangle \big] \\ & = \sum_k e^{i \delta_k t} (\langle X_k \rangle + i \langle Y_k\rangle),
\end{split}
\end{equation}
where it is implied that the expectation value $\langle \rangle$ is performed over the initial ansatz $|\Psi_0\rangle$. We have explicitly decoupled the time evolution of the magnetization operator, involving only phases of the chemical shifts in the NMR Hamiltonian. 

We can now compute the Fourier transform $\mathcal{F}$ of the FID signal, the magnetization signal attenuated with a damping factor $e^{-\eta t}$, by using the following integral relation 
\begin{equation}\label{eq:int_rel}
    \int_0^{\infty} dt e^{-\eta t} e^{-i (a - b) t} = \frac{1}{\eta + i (a -b)}.
\end{equation}
Finally,
\begin{equation}
\begin{split}
    \textrm{NMR} &= \textrm{Re}\mathcal{F}[e^{-\eta t} \langle \Psi(t)|M|\Psi(t)\rangle ] \\& = \frac{1}{2}\textrm{Re}\sum_k (\langle X_k \rangle + i \langle Y_k\rangle)\int_0^{\infty} dt e^{-\eta t} e^{i\delta_kt}e^{-i 2 \pi f t} \\ &= \frac{1}{2}\textrm{Re}\sum_k \frac{\langle X_k \rangle + i \langle Y_k\rangle}{\eta + i (2 \pi f - \delta_k)} \\& =\frac{1}{2}\sum_k \frac{\eta \langle X_k\rangle + (2\pi f - \delta_k)\langle Y_k \rangle }{\eta^2 + (2\pi f - \delta_k)^2}, 
\end{split}
\end{equation}
which, for a Hadamard state as initial ansatz, $\langle X \rangle = 1$ and $\langle Y \rangle = 0 $, and the expression reduces to the standard Lorentzian peaks centered at the chemical shifts. 

\section{Peak splitting}
\begin{figure}
    \centering
    \includegraphics[width=0.95\columnwidth]{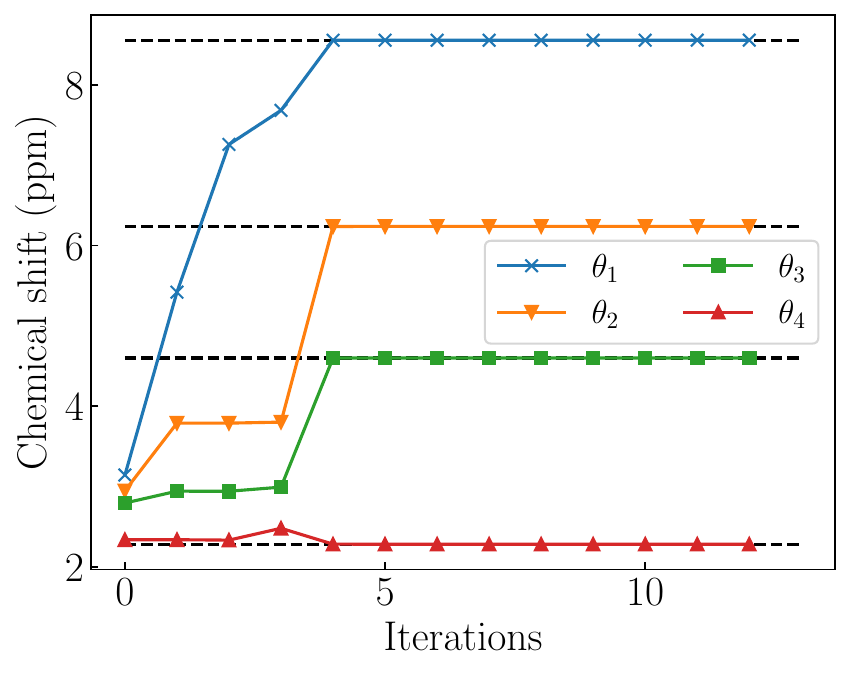}
    \caption{Convergence of peak positions $\theta_k$ in the cost function with the iterative MM-QCELS pipeline for sulfanol and using only a $ZZ$ interaction in the NMR Hamiltonian. The dashed lines correspond to the exact values obtained as explained in the text.}
    \label{fig:sulfanol_only_ZZ}
\end{figure}

In this Section, we give a back of the napkin derivation of splitting seen on the peaks of the NMR spectrum. If we consider a NMR Hamiltonian of the form 
\begin{equation}
    H = \frac{1}{2}\sum_i \delta_i Z_i + \frac{1}{4}\sum_{i<j} J_{ij} Z_iZ_j,
\end{equation}
we can perform the exact time evolution of the magnetization signal, owing to the following expressions
\begin{equation}
\begin{split}
    e^{i \theta Z_j Z_k} X_k e^{-i \theta Z_jZ_k} &= \cos(2\theta) X_k - \sin(2\theta) Z_j Y_k, \\
    e^{i \theta Z_j Z_k} Y_k e^{-i \theta Z_j Z_k} &=  \cos(2\theta) Y_k + \sin(2\theta) Z_j X_k, \\ &
    \textrm{for    } \,\, \,\, j \neq k.
\end{split}
\end{equation}
Without loss of generality, let us consider a system with two spins
\begin{equation}
    H = \frac{\delta_0}{2} Z_0 + \frac{\delta_1}{2} Z_1 + \frac{J_{01}}{4} Z_0 Z_1,
\end{equation}
whose evolution operator can be expressed as independent products 
\begin{equation}
    e^{-i H t } = e^{- i \frac{\delta_0}{2} Z_0} e^{- i \frac{\delta_1}{2} Z_1} e^{- i \frac{J_{01}}{4} Z_0 Z_1}.  
\end{equation}
For the $X_0$ operator we have 
\begin{equation}
\begin{split}
    e^{i H t } X_0 e^{-i H t } &= 
    e^{i \frac{J_{01}}{4} Z_0 Z_1 t} e^{i \frac{\delta_0}{2} Z_0 t} X_0 e^{-i \frac{\delta_0}{2} Z_0 t} e^{- i \frac{J_{01}}{4} Z_0 Z_1 t}\\&
    = e^{i \frac{J_{01}}{4} Z_0 Z_1 t}  [\cos(\delta_0 t) X_0 - \sin(\delta_0 t) Y_0]e^{-i \frac{J_{01}}{4} Z_0 Z_1}\\&
    = \cos \left( \frac{J_{01}}{2} t\right) [\cos(\delta_0 t) X_0 - \sin(\delta_0 t) Y_0] - \\& - \sin \left( \frac{J_{01}}{2} t\right) Z_1 [\cos(\delta_0 t) Y_0 + \sin(\delta_0 t) X_0] ,
\end{split}
\end{equation}
and, equivalently for the $Y_0$ operator 
\begin{equation}
\begin{split}
    e^{i H t } Y_0 e^{-i H t } &= 
    e^{i \frac{J_{01}}{4} Z_0 Z_1 t} e^{i \frac{\delta_0}{2} Z_0 t} Y_0 e^{-i \frac{\delta_0}{2} Z_0 t} e^{- i \frac{J_{01}}{4} Z_0 Z_1 t}\\&
    = e^{i \frac{J_{01}}{4} Z_0 Z_1 t}  [\cos(\delta_0 t) Y_0 + \sin(\delta_0 t) X_0] \times \\& \times e^{-i \frac{J_{01}}{4} Z_0 Z_1 t}\\&
    = \cos \left( \frac{J_{01}}{2} t\right) [\cos(\delta_0 t) Y_0 + \sin(\delta_0 t) X_0] - \\& - \sin \left( \frac{J_{01}}{2} t\right) Z_1 [\sin(\delta_0 t) Y_0 - \cos(\delta_0 t) X_0].
\end{split}
\end{equation}
So, for the whole magnetization $M_0$ on the first spin we have, using the Hadamard state as the reference ansatz, 
\begin{equation}
\begin{split}
    \langle \Psi(t)|M_0|\Psi(t)\rangle &= \frac{1}{2} \langle \Psi(t)|(X_0 + iY_0)|\Psi(t)\rangle \\&= \frac{1}{2} \cos \left( \frac{J_{01}}{2} t\right) [\cos(\delta_0 t) + i \sin(\delta_0 t) ] \\&= \frac{1}{2}  \cos \left( \frac{J_{01}}{2} t\right) e^{i \delta_0 t}
\end{split}
\end{equation}
Thus, for the whole magnetization we finally obtain 
\begin{widetext}
\begin{equation}
\begin{split}
    \langle \Psi(t)|M|\Psi(t)\rangle &= \frac{1}{2} \langle \Psi(t)|M_0 + M_1|\Psi(t)\rangle = \frac{1}{2}  \cos \left( \frac{J_{01}}{2} t\right) \left(e^{i \delta_0 t} + e^{i \delta_1 t}\right) = \frac{1}{4} \left(e^{-i \frac{J_{01}}{2} t} + e^{i \frac{J_{01}}{2} t}\right) \left(e^{i \delta_0 t} + e^{i \delta_1 t}\right)  \\& = \frac{1}{4} \bigg[e^{i\left(\delta_0 - \frac{J_{01}}{2}\right)t} + e^{i\left(\delta_0 + \frac{J_{01}}{2}\right)t} + e^{i\left(\delta_1 - \frac{J_{01}}{2}\right)t} + e^{i\left(\delta_1 + \frac{J_{01}}{2}\right)t} \bigg].
\end{split}
\end{equation}
\end{widetext}
From this expression, we can see that in the frequency domain after a Fourier transform using again the relation in Eq.~(\ref{eq:int_rel}), we will observe four peaks at $\delta_0 \pm \frac{J_{01}}{2}$ and $\delta_1 \pm \frac{J_{01}}{2}$, corresponding to a splitting of $\frac{J_{01}}{2}$ on the peaks corresponding to spins $0$ and $1$. The introduction of more spin couplings, between spin $a$ and $b$, will further split the peaks with additional an additional $\pm \frac{J_{ab}}{2}$ factor. In fact, the same splitting will be seen when considering the full NMR Hamiltonian, as the additional $XX$ and $YY$ terms commute with $ZZ$, and only affect the position of the peak centers. This can be seen in Fig.~\ref{fig:sulfanol_only_ZZ}, where we run a MM-QCELS simulation with only a $ZZ$ interaction in the NMR Hamiltonian for sulfanol. We see the convergence towards the exact values given by the formulas mentioned before and the parameters shown in the Results section of the main text.


\bibliography{references}

\end{document}